# The equation of state of a partly homogenized plasma of low-dense porous matter

S.Yu. Gus'kov, R.A. Yakhin

P.N. Lebedev Physical Institute of Russian Academy of Sciences. Leninskiy Av., 53, Moscow 119991, Russia

E-mail: yakhin.rafael@gmail.com

**Abstract**

The equation of state (EOS) of a low-density porous substance plasma is proposed in the form of continuous media EOS containing, as a pressure control parameter, the degree of plasma homogenization, which is function of the initial porous structure, as well as of the current values of plasma density and temperature. Using the partially-homogenized-plasma EOS an approximate analytical solution is found and numerical calculations were performed of the problem of thermal expansion of a flat layer of porous matter. The features of the obtained results are discussed in comparison with the case of a continuous substance of equivalent chemical composition. The proposed equation of state is used to analyze the experimental data on porous substance heating with laser and X-ray pulses.

## I. Introduction

Investigation of the properties of plasma produced when a low-density porous substance is heated by powerful pulses of laser or X-ray radiation is an actual section of modern high energy density physics. This is due to both the fundamental nature of physical processes underlying the formation of such a plasma and the important significance of associated applications. The main feature of porous substance plasma is homogenization that is leveling of its density inside the pore /1/. This process determines the properties of non-stationary and non-equilibrium state of such a plasma. The rate of homogenization depends on the porous structure parameters as well as on the power of heating source. It is important that some fraction of the internal energy of partially homogenized plasma is related to the energy of turbulent motion of plasma flows inside the pores. This is the reason for slow formation of directed hydrodynamic motion and electron heat conduction wave during the homogenization period /1, 2/. Micro-size porous media of light chemical elements, such as porous plastics ($CH_n$), agar-agar ($C_{12}H_{18}O_9$), triatsetat-cellulosa ($C_{12}H_{16}O_8$), trimethylolpropane-triacrylate ($C_{15}H_{20}O_6$), and polyvinylalcohol ($[CH_2CH(OH)]_n$) with a density from several units to several tens of mg·cm$^{-3}$ /3/ are of the greatest interest for research. The porosity of these media, i.e. the relative share of void's volume, is 0.99-0.999 /3/. Their structure can be interconnected quasi-flat membranes or extended filaments, but, most often, it is a mixed membrane-filament structure. Structural parameters can vary over a wide range for different substances. The average pore's size can range from several tens of microns to submicron values with the average thicknesses of solid elements (membranes or filaments) of micron and submicron scale, respectively. These media withstand the additions of nanoscale clusters of heavy elements such as gold or copper with weight fraction of 20-30% /3/.

One of the directions of studying a laser-produced plasma of porous media is associated with their use as an absorber of high-power laser radiation, which is capable to provide a large pressure of the resulting plasma at effective alignment of its heating nonuniformities under the conditions of experiments on laser thermonuclear fusion (ICF) /4, 5, 6/ and equation of state (EOS) /7-10/. This is due to a volumetric nature of laser radiation absorption at the depth of geometric transparency of porous media with an average density, both lower and higher than the critical plasma density, corresponding to the wavelength of acting radiation. A large fraction (80–90%) of the 1st–3rd Nd laser harmonic radiation absorbed in low-dense porous media, which is actually independent of these harmonic wavelength, has been established in many experiments /11–16/. Another direction of research is related to creating the laser or X-ray produced plasmas with controlled temporal and spatial evolution of thermodynamic parameters. In such plasmas, for example, generation of high-energy charged particles under the action of a petawatt laser pulse /17-19/ or slowing down of ion or electron beams /20, 21/ are actively investigated. At the same time, unlike a gaseous medium, the use of a porous substance as an absorber of laser or soft X-ray radiation in the targets for various purposes does not require special technical efforts.

The model of forming a partially homogenized plasma of a micro-size porous substance of light elements as a result of collisions between the flows of evaporated substance of pore's walls was proposed in /1, 2, 12, 22/ and developed in /23-30/. In particular, in /2, 22/ the dependence of homogenization degree, as a relative fraction of homogenized substance mass, on temperature, as well as on average density and initial porous structure parameters was found. Numerical hydrodynamic calculations using the partially-homogenized-plasma model showed quantitative agreement with the results of experiments, including the slower propagation of laser-produced shock wave in a porous substance compared to the case of a continuous substance of equivalent chemical composition /23-33/.

This paper is devoted to further development of theory of partially homogenized plasma, namely, to creation of such a plasma EOS in the form of continuous medium EOS with pressure controlling parameter which is plasma homogenization degree defined in the above-mentioned papers /2, 22/. The EOS model is presented in Section II. Section III is devoted to using the partially-homogenized-plasma EOS to describe a thermal expansion of uniformly heated layer of porous matter.

## II. The equation of state of partly homogenized plasma

A partly homogenized plasma of a porous substance can be represented as a two-component medium with fractions of homogenized and non-homogenized plasmas. The first fraction of plasma homogenized by a given moment of time consists of the ions and electrons with number corresponding to the ion's number from the electrical neutrality condition, which perform a translational thermal motion and thereby give the contribution to pressure. The non-homogenized fraction consists of the ions and the corresponding number of electrons, which by a given moment of time remain involved in a turbulent motion of plasma flows colliding inside the pores and do not contribute to pressure. The mass ratio of homogenized and non-homogenized fractions is determined by homogenization rate, which is rate of diffusion broadening the homogenized plasma region as a result of ion's Coulomb collisions in plasma flows interacting inside the pores /2, 22/. In order of magnitude, the homogenization rate is the ratio of ion-ion collision length to the time of passage through a pore of plasma flow from heated pore's wall. This rate depends on the parameters of porous structure and the power of heating source /22/. To construct such a two-component plasma EOS in the form of a continuous medium EOS, it is necessary to introduce the partial density $\rho_h$ of homogenized fraction through the homogenization degree H, that is through the relative fraction of homogenized fraction mass per unit volume at the given point with density $\rho$ and temperature T: $\rho_h(x,t)=H(x,t)\rho(x,t)$.

The homogenization degree H can only be set approximately, as the homogenization degree of an individual pore with an average density $\rho$ at the heating temperature of its walls T. This approximation gives the more accurate result, the smaller the pore size compared to the scalelengths of thermodynamic parameters. In view of a low density of considered porous media, the plasma of homogenized fraction and the plasma of flows colliding inside the pores are assumed to be ideal. In the one-temperature approximation, in the first case, the internal energy is determined by the energy of thermal motion of particles with temperature T, and in the second one, by the kinetic energy $T/(\gamma_c-1)$, so that the total internal energy $\varepsilon$ is determined by the total density $\rho$: $\varepsilon=C_V\rho T$, where $C_V=(Z+1)k_B/[(\gamma_c-1)Am_p]$ and $\gamma_c$ are the specific heat and adiabatic index of a completely homogenized plasma; Z and A are the charge and atomic weight of plasma ions, $m_p$ is the proton mass, $k_B$ is the Boltzmann constant. Then, dividing the internal energy $\varepsilon$ into two components related to the homogenized and non-homogenized fractions, in accordance with homogenization degree H, the pressure of partially homogenized plasma in the average-ion approximation ($\gamma_c=5/3$) is written as:

$$P = P_c(x,t)H(x,t), \qquad (1)$$

where $P_c=C_V(\gamma_c-1)\rho T$ is pressure of a continuous plasma, in which the density is equal to the average density of porous substance. In terms of pressure and internal energy, the equation of state (1) has the usual form

$$P = (\gamma-1)\varepsilon, \qquad (2)$$

where, however, the partially-homogenized-plasma adiabatic index $\gamma$ is related to the adiabatic index of continuous plasma, as

$$\gamma(x,t) = 1 + (\gamma_c - 1)H(x,t). \qquad (3)$$

In non-homogenized plasma at H=0 there is $\gamma$=1 and P=0 i.e. energy is contained only in the turbulent motion of plasma flows inside the pores with an infinite number of freedom's degrees. When H=1, the plasma is completely homogenized, hence, $\gamma=\gamma_c$ and pressure is equal to the total pressure $P=P_c$ of all particles, which all perform a translational motion in this case.

In /22/, based on the calculation of diffusion broadening of homogenized plasma region during multiple interactions of plane plasma flows from the pore's walls located opposite each other with a thickness much smaller than pore's size, the time dependence of homogenization degree of an individual pore was obtained in the form

$$H(x,t) = 1 - \left[1 - 2\int_0^t \frac{dt'}{t_p(x,t')}\right]^{1/2}, \qquad (4)$$

where $t_p$ is the of characteristic homogenization time, which for a pore with initial size $\delta_0$ and average density $\rho_0$ at the heating temperature T has the form

$$t_{p0}[s] = \frac{\delta_0^2}{V_i^2 \tau_{ii}} \approx 2.4 \cdot 10^{-11} \frac{Z^4 \delta_0^2 \rho_0}{A^{1/2} T^{5/2}}, \qquad (5)$$

$V_i$ is the average ion's velocity of plasma flows inside a pore, $\tau_{ii}$ is the time of ion-ion collisions. The expression (5) for homogenization degree of an individual pore can be applied to approximately determine the partial density $\rho_h(x,t)$ using the normalization based on the condition that the mass of a unit volume of continuous medium is equal to the pore's mass $\rho_0\delta_0^3$. This normalization gives the correction to expression (5) in the form of weak dependence on the density of continuous medium $\rho$

$$t_p = t_{p0}\left(\frac{\rho}{\rho_0}\right)^{1/3}. \quad (6)$$

The homogenization degree (4) calculated using the characteristic time in the form (5) or in the refined form (6), is a necessary element of partially-homogenized- plasma EOS.

### III. Thermal expansion of porous matter layer

Let us apply the partially-homogenized-plasma EOS to the problem of thermal expansion of a flat layer of a porous substance with a given thickness L instantly heated to a temperature T. The problem is interesting from the point of view of controlling the temporal and spatial evolutions of thermodynamic parameters of plasma produced by laser or X-ray heating of a porous substance. The analytical solution obtained below and the numerical calculations performed clearly demonstrate the nontrivial nature of partially-homogenized-plasma EOS effect. The duration of layer's thermal expansion is determined by the duration of convergence in the layer's center of the fronts of rarefaction waves propagating inside the layer from its edges. This duration will considered as a time scale during which quasi-consolidated state of layer is, approximately, conserved. The temporal dependence of reduction in thickness of the layer's central part, which is not covered by motion, is determined by velocity of rarefaction wave, which in turn depends on the pressure in this very central part of layer. Moreover, the pressure in layer's central part, decreasing in thickness, changes with time due to homogenization occurring in it. The pressure increases from 0 at the initial moment of time, when the porous substance is not homogenized, to a certain maximum value, which is determined by the ratio of homogenization and expansion durations. This distinguishes the expansions of a porous substance layer (the term

"porous layer" will be briefly use below) and equivalent layer of continuous matter ("continuous layer"), in which rarefaction wave velocity is equal to sound velocity independent on time and corresponding just to the heating temperature T.

The homogenization degree in central region of porous layer, not covered by expansion, at a constant temperature and density, according to (1) and (2), is

$$H = \begin{cases} \left[1 - \left[1 - \frac{t}{t_h}\right]^{1/2}\right]^{1/2}, & t \leq t_h \\ 1, & t \geq t_h \end{cases} \quad (7)$$

where the homogenization time $t_h = t_p/2$. Then, according to expression (1) with taking into account (7), pressure and sound velocity in a partially homogenized plasma are calculated as

$$P = P_c \left(1 - (1-\tau)^{1/2}\right), \quad V = V_c \left[1 - (1-\tau)^{1/2}\right]^{1/2}, \quad \tau \leq 1 \quad (8)$$

here $\tau = t/t_h$ and $P_c$ and $V_c$ are pressure and sound velocity in an equivalent continuous plasma.

Therefore, in a thick porous layer where homogenization's duration is less than expansion's one, pressure in central part of layer not covered by rarefaction wave and this wave velocity increase from zero at the initial moment of time to, respectively, pressure and sound velocity of continuous plasma, which are reached at the end of homogenization process $t=t_h$ ($\tau=1$). The lower velocity of rarefaction wave during the period of homogenization leads to increase in the duration of maintaining the layer in a quasi-consolidated state. The use of second expression of (8) leads to simple approximate solution for the total width of two edge regions of porous layer covered by the rarefaction waves

$$\Delta_r = \Delta_{rc} \begin{cases} \frac{2}{5} \tau^{1/2}, & \tau \leq 1 \\ 1 - \frac{3}{5\tau}, & \tau \geq 1 \end{cases}, \quad (9)$$

here $\Delta_{rc} = 2V_c \cdot t$ is the total width of regions covered by rarefaction waves in equivalent continuous layer. Using (9), it is easy to obtain the duration of maintaining a quasi-consolidated state of porous layer with a given thickness L

$$t_r = t_{rc} \begin{cases} 1 + \dfrac{3}{5} \cdot \dfrac{t_h}{t_{rc}}, & 0 \leq \dfrac{t_h}{t_{rc}} \leq \dfrac{5}{2} \\ \left(\dfrac{5}{2}\right)^{2/3} \left(\dfrac{t_h}{t_{rc}}\right)^{1/3}, & \dfrac{t_h}{t_{rc}} \geq \dfrac{5}{2} \end{cases}, \quad (10)$$

here $t_r$ is the time of rarefaction wave propagation in porous layer over a distance equal to half the layer thickness L/2, $t_{rc}=L/2V_c$ is the same time in the case of equivalent continuous layer

$$t_{rc}[s] \approx 1.3 \cdot 10^{-8} \cdot \left[\dfrac{A}{(Z+1)}\right]^{1/2} \dfrac{L}{T^{1/2}}. \quad (11)$$

Finally, substituting (10) into expression (4) makes it possible to determine the pressure scale in porous layer, which takes place during the existence of a quasi-consolidated state of layer

$$P = P_c \begin{cases} 1, & 0 \leq \dfrac{t_h}{t_{rc}} \leq \dfrac{5}{2} \\ \left[1 - \left(1 - \dfrac{5}{2} \cdot \dfrac{t_{rc}}{t_h}\right)^{2/3}\right]^{1/2}, & \dfrac{t_h}{t_{rc}} \geq \dfrac{5}{2} \end{cases}, \quad (12)$$

where $P_c$ is initial pressure in continuous layer.

Both the ratios $t_r/t_{rc}$ and $P/P_c$ are functions of the only dimensionless parameter of the problem, which is the ratio of the homogenization time $t_h$ to the time of rarefaction wave front propagation over half the thickness of equivalent continuous layer $t_{rc}$. Using (5) and (11), this parameter is written as

$$\dfrac{t_h}{t_{rc}} \approx 9.2 \cdot 10^{-4} \dfrac{Z^4 (Z+1)^{1/2} \delta_0^2 \rho_0}{2 A^{3/2} T^2 L}. \quad (13)$$

The dependences of $t_r/t_{rc}$ and $P/P_c$ values on $t_h/t_{rc}$ parameter are shown in Fig. 1.

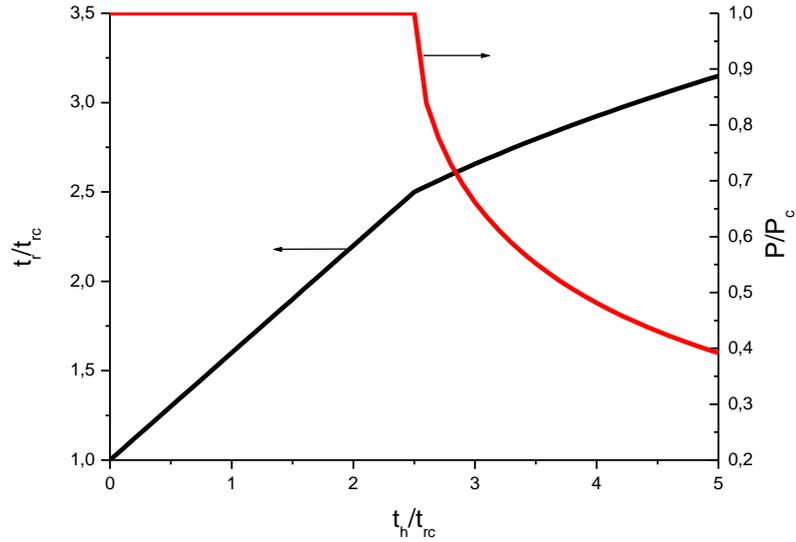

Fig. 1. Relationships between the durations of a quasi-consolidated state existence (black line) and the pressures (red line) in this state in porous and equivalent continuous layers on the ratio of homogenization time to the duration of rarefaction wave propagation over half the thickness of continuous layer $t_h/t_{rc}$.

When the homogenization time tends to 0, i.e. when ratio $t_h/t_{rc}$ tends to 0, the duration of a quasi-consolidated state existence and pressure scale tends to the corresponding values of continuous layer. With an increase in the homogenization time $t_h$ compared to $t_{rc}$, the excess of duration of quasi-consolidared state existence increases compared to the case of a continuous layer with a decrease in the pressure scale compared to the pressure scale of a continuous layer. In other words, the larger $t_h/t_{rc}$ parameter, i.e. the longer homogenization takes place compared to the time of layer thermal expansion, the greater partly-homogenized-plasma EOS effects in a porous layer, and vice versa. Note that parameter $t_h/t_{rc}$ strongly depends on temperature, like $t_h/t_{rc} \propto T_0^{-2}$, due to the strong, according to (5), homogenization time dependence on temperature. Therefore, the excess of the duration of quasi-consolidated state existence of a porous layer in comparison with a continuous layer increases with increase in heating temperature.

Let us apply the partly-homogenized-plasma EOS for interpreting the results of experiments on porous substance heating by laser and soft X-rays radiations, in which the porous substances and the conditions of their heating were strongly differed. In addition to using the analytical solution described above, the numerical calculations of uniformly heated porous layer were performed with help the one-dimensional hydrodynamic code DIANA /34/ with the partly-homogenized-plasma EOS and the ideal gas EOS. The calculations were performed with electron thermal conductivity turned off in order to maximize EOS effect. As an example of experiments on laser heating, the experiments /33/ were chosen, which are interesting in that the homogenization duration was established there. In these experiments, which were carried out on ABC laser (ENEA-Frascati, Italy), the layers of large-pore polystyrene of 400-500 μm thick were irradiated with the 1-st Nd-laser harmonic radiation pulse with an energy of about 30 J, duration of about 3 ns at the intensity on a layer's surface of about $5 \cdot 10^{13}$ W·cm$^{-2}$. The average density of porous polystyrene was about ρ≈10 mg·cm$^{-3}$. Average pore's size and wall thickness are of about $\delta_0$≈40 μm and 1 μm, respectively. The initial solid density of pore's wall was about $\rho_s \approx 1$ g·cm$^{-3}$. The substance had a mixed membrane-filaments structure, in which the length of geometric transparency that is the scale of laser radiation absorption depth, according to /22/, is calculated as

$$L_g [cm] \approx 5 \cdot 10^{-4} \cdot \left(\frac{\rho_s}{\rho}\right)^{0.2} \delta_0. \quad (14)$$

For the above mentioned porous substance parameters, the geometric transparency length is $L_g \approx$ 500 μm. The fact that the thicknesses of layers used in the experiments under discussion were close to geometric transparency length allows us to assume that the porous layers were, approximately, uniformly heated. The homogenization duration of about 2.5 ns was determined from the termination of scattered laser radiation oscillations. According to (5), this value for a porous substance with the indicated parameters corresponds to a heating temperature of 1.8 keV /33/. Further calculation using the formulas (11) and (13) at the temperature T =1.8 keV and the layer thickness L=$L_g$=500 μm gives the values of about 0.6 ns, and 4.8, respectively, for time $t_{rc}$ and $t_h/t_{rc}$ parameter, which means a significant effect of homogenization on thermal expansion rate of porous layer.

According to (10), at $t_h/t_{rc}$=4.8, the duration of quasi-consolidated state maintaining is about $t_r$=1.8 ns that is almost 3 times longer than in the case of equivalent continuous layer. In turn, according to (12), by the end of the period of quasi-consolidated state existence the pressure reaches only 43% of the pressure in equivalent continuous layer. Numerical calculations performed at L=500 μm and T=1.8 keV confirm the results of the analytical solution. The time for rarefaction wave to reach the middle of the porous layer was $t_r$=1.3 ns, which is 2.6 times longer than the time for the wave to reach the middle of equivalent continuous layer (0.5 ns). By the time $t_r$=1.3 ns, the pressure in the middle of layer was 5.75 Mbar, which was 48% of the pressure of equivalent continuous substance heated to the temperature of 1.8 keV (12 Mbar). As already mentioned in the introduction, in various experiments on laser irradiation of porous layers with a thicknesses exceeding the geometric transparency length, including the experiments /25, 28, 30, 31, 32/, a decrease in a shock wave velocity was established compared to the case of equivalent continuous layer. This decrease in various experiments is 1.2-1.7 times in dependence on the intensity of laser pulse and the characteristics of porous substance. In particular, in the experiments /32/ performed under the same conditions as the experiments /33/, this decrease was about 1.5 times. This indicates that the proposed equation of state, which gives the decrease in pressure by a factor of 2.5 and the decrease in sound velocity by a factor of 1.6 under experimental conditions /32, 33/, correctly describes the experimental results.

As experiment on X-ray heating of a porous layer, let us consider the experiment /20/, which present the data on quasi-consolidated state maintaining of produced plasma. In these experiments, the layer of a finely porous TAC substance with pore's size of about 1 μm, average density of 2 mg·cm$^{-3}$ and thickness of 1000 μm was heated by soft X-ray radiation to the temperature of about 17 eV. It was reported that the heated layer maintained a quasi-consolidated state for the time period of more than 10 ns. The parameters of porous structure and the heating temperature were very different from the example of experiments on laser heating considered above. The density of porous substance was 5 times less, and the pore's size was 40 times greater. The layer thickness was about 2 times greater. Finally, the heating temperature was two orders of magnitude lower. According to the formulas (5) and (11) for the conditions of X-ray experiment at the average ion's charge of Z=2.5, specified in /20/, the homogenization time is $t_h \approx 9.5$ ns, the time $t_{rc}$ is about 11.2 ns and, accordingly, the parameter $t_h/t_{rc}$ is 0.9. Then, according to (10), for the time of quasi-homogeneous state maintaining, we obtain $t_r \approx 16.8$ ns, which is in good agreement with the experimental data. In contrast to the case of laser heating considered above, in which homogenization proceeded much more slowly compared to the time of a continuous layer expansion and in which the ratio $t_h/t_{rc}$ was 4.8, in the case of X-ray heating, the homogenization time is shorter than the time of continuous layer expansion. This means that the role of homogenization in delaying the expansion of porous layer is small and leads to an increase in the expansion time by only a factor of 1.4, while in the case of laser heating experiment, in which the role of homogenization in delaying the expansion is large, it led to an increase in the expansion time by factor of 3. In numerical calculations performed at L=1000 μm and T=17 eV, the time for rarefaction wave to reach the middle of porous layer was $t_r$=16 ns, which is only 1.2 times longer than the time for the wave to reach the middle of equivalent continuous layer (12.5 ns). By the time $t_r$=1.2 ns, the pressure in the middle of layer was 0.017 Mbar, which is close to the pressure of continuous substance ionized to Z=2.5 at the temperature of 17 eV (0.02 Mbar).

To illustrate the partially-homogenized-plasma EOS effect, Figs. 2-5 show some additional results of the above-mentioned numerical calculations of thermal expansion of laser-heated porous polystyrene layer (ρ=10 mg·cm$^{-3}$, $δ_0$=40 μm, L=500 μm, T =1.8 keV) under the conditions of experiments /33/. For comparison, the results of calculation of thermal expansion of equivalent layer of continuous polystyrene-"gas" (ρ=10 mg·cm$^{-3}$, L=500 μm) also are shown. Fig. 2 shows the spatial distributions of pressure, density, and temperature, as well as the homogenization degree at the different time moments of a porous layer expansion when using the partly-homogenized-plasma EOS with the normalized homogenization time (6). Fig. 3 shows the spatial distributions of pressure, density and temperature at the different time moments of an equivalent continuous layer expansion with an ideal gas EOS. Fig. 4 shows the time dependences of relative fraction of the energy of hydrodynamic motion in expanding layers of porous and continuous matter.

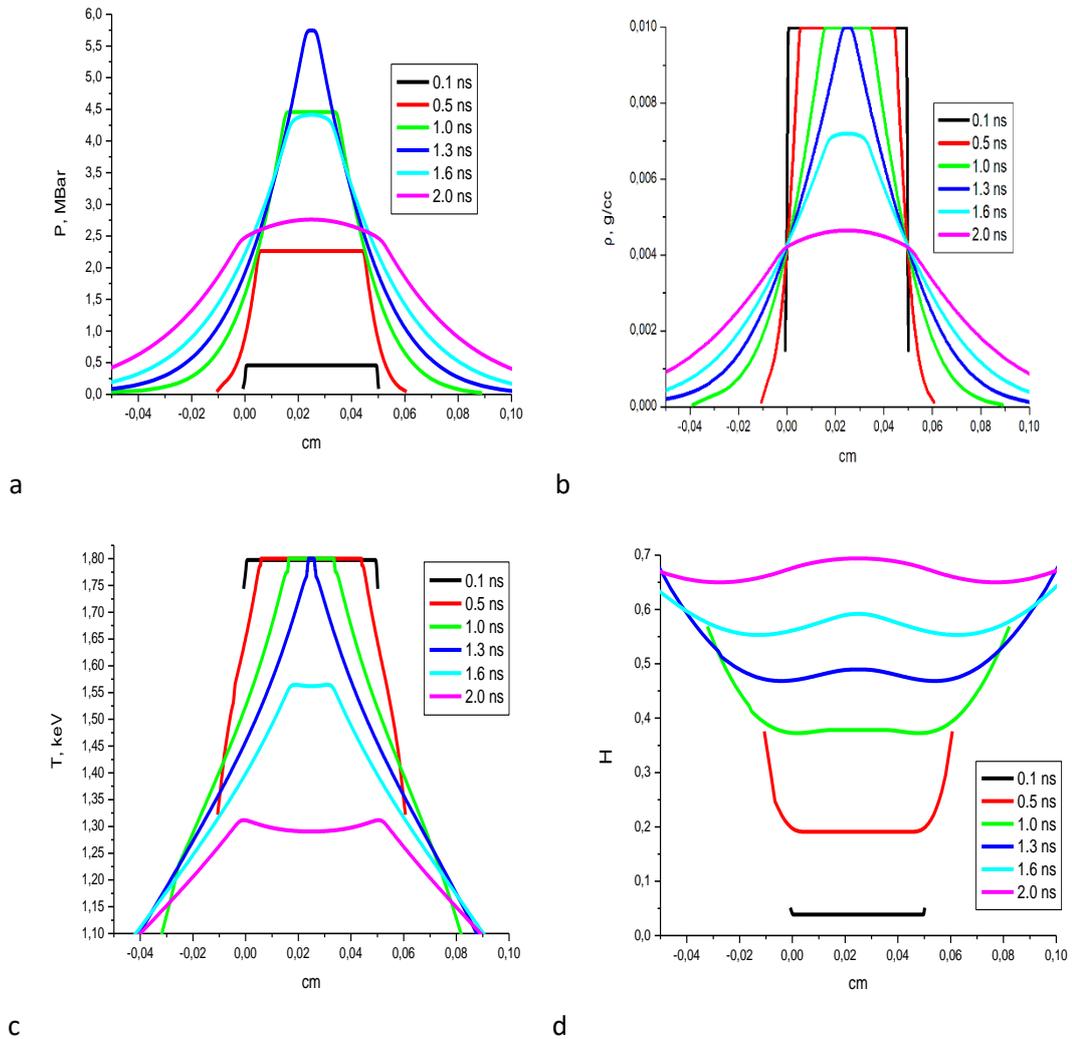

Fig. 2. Spatial distributions of pressure (a), density (b), temperature (c), and degree of homogenization (d) at the different time instants of a porous layer expansion with the partly-homogenized-plasma EOS.

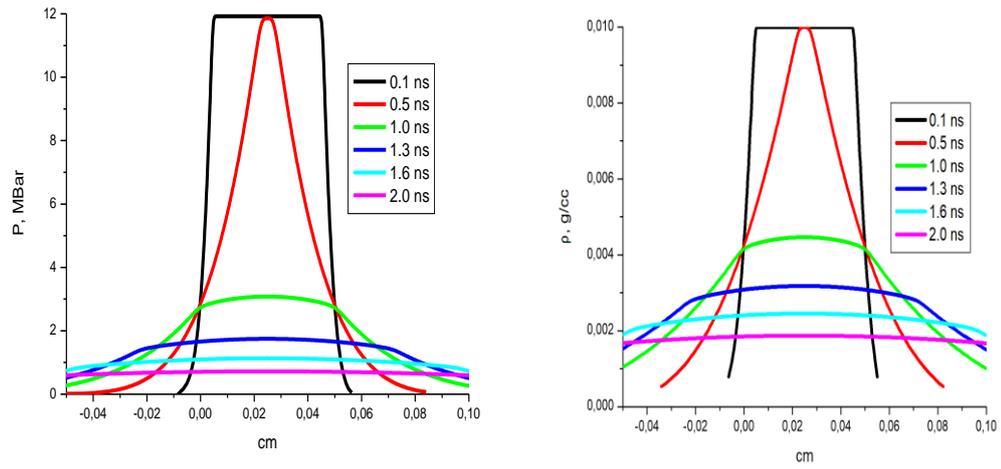

Fig. 3. Spatial distributions of pressure (a), density (b), and temperature (c) at the different time instances of a continuous layer expansion with the ideal gas EOS.

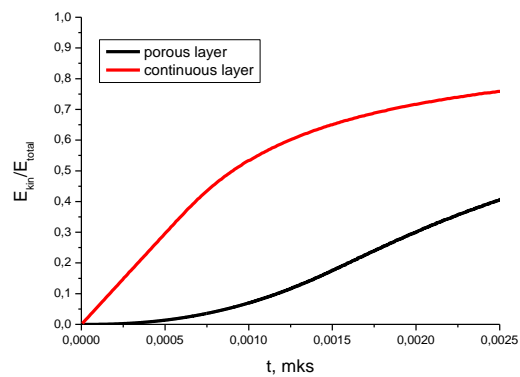

Fig. 4. Time dependences of relative fraction of the energy of hydrodynamic motion during the expansion of porous layer (black line) and equivalent continuous layer (red line).

The calculations clearly show the above-mentioned main feature of partly-homogenized-plasma EOS effect, which manifests itself in pressure temporal evolution, which is opposite to the case of continuous layer. In the case of continuous layer, the pressure in plasma regions covered by motion, naturally, decreases monotonically (Fig. 3a). In the case of porous layer, with an increase in homogenization degree with time (Fig. 2 d), the pressure, on the contrary, increases from zero, corresponding to the initial state of non-homogenized substance, up to a maximal value in the moment t≈1.3 ns of beginning the layer motion as a whole, and only after that it starts to decrease (Fig. 2 a). The maximal pressure of partially homogenized plasma at the time $t_r$≈1.3 ns is 5.75 Mbar. These value are only 7–8 % larger than the values of analytical solution presented above. By the time of 1.3 ns, the homogenization degree averaged over expanding layer thickness is about 0.54. The spatial distribution of this value has a local maximum of 0.47 in the middle of layer, where the dependence of homogenization time on temperature is most pronounced, and maximum values of about 0.62 at the edges of layer, where the dependence of homogenization time on density is most pronounced (Fig. 2d). As time increases, the homogenization degree distribution levels off over the layer thickness. By the time of 2 ns, homogenization degree distribution is close to uniform one with average value of about 0.67. A local temperature minimum in the middle of layer is associated with a local maximum of homogenization degree, since a larger homogenization degree means a larger rate of the thermal energy transformation into the energy of hydrodynamic motion. The beginning of motion of continuous layer as a whole occurs at the time moment t=0.5 ns much earlier than in the case of porous layer (1.3 ns). By the time of 1.3 ns, the pressure in the middle of continuous layer decreases by a factor of 6.7 compared to the initial value and is only 1.8 Mbar (Fig. 3a).

Thermal expansion of a porous layer occurs much more slowly than expansion of an equivalent continuous layer. At the time moment 1.3 ns, when the initial values of density (0.01 g·cm$^{-3}$) and temperature (1.8 keV) remain in the middle of porous layer (Fig. 2b and 2c), the values of these thermodynamic parameters in the middle of continuous layer are 0.003 g·cm$^{-3}$ and 0.8 keV (Fig. 3b and 3c). At the time of 2 ns in the middle of porous layer, the values of density and temperature are, respectively, 0.0043 g·cm$^{-3}$ and 1.27 keV and in the middle of continuous layer they are 0.0017 g·cm$^{-3}$ and 0.55 keV. A much slower thermal energy transition into hydrodynamic motion energy during the expansion of porous layer compared to the expansion of continuous layer is illustrated in Fig. 4. By the time of 1.3 ns, the relative fraction of hydrodynamic motion energy is 0.12 in a porous layer and 0.52 in continuous layer. At the time instant of 2 ns, these values are 0.27 and 0.63, respectively.

Numerical calculations with the partially-homogenized-plasma EOS with the simplified characteristic homogenization time in the form (5), which does not depend on density, showed results that are practically indistinguishable from those presented above in Figs. 2a-2c over the entire region of layer, except for the temperature distribution in the regions of low-density matter at layer's edges. Due to the lower homogenization degree in the case of EOS with homogenization time (5) and the associated lower thermal energy transformation into energy of hydrodynamic motion, the temperature at layer's edges was approximately 20-30% larger compared to the calculations with the normalized homogenization time (6). The insignificant difference the results of both calculations is illustrated by the fact that the relative fraction of hydrodynamic motion energy in the case of EOS with homogenization time (5) by the time 2 ns is 0.28 that is only 2 % lower than this value in the case of normalized homogenization time (Fig. 4).

## Acknowledgements

The research was financially supported by Russian Science Foundation under the Project No. 21-11-00102.

## Conclusion

An equation of state for partly homogenized plasma of a low-density porous substance in the form of EOS of a continuous medium is proposed. It contains, as a control parameter of pressure, the degree of plasma homogenization, which is function of the parameters of initial porous substance as well as of density and temperature of continuous medium. An analytical solution for thermal expansion of a porous layer with partially-homogenized-plasma EOS is obtained, which is in good quantitative agreement with the results of numerical calculations. It is shown that the partially-homogenized-plasma EOS effect on thermodynamic and hydrodynamic characteristics is determined by a dimensionless parameter, which is the ratio of the homogenization duration to the duration of rarefaction wave propagation over equivalent layer of continuous substance. The dependence on this parameter of the degree of increase in the duration of maintaining the quasi-consolidated state of a heated layer of porous substance in comparison with the case of equivalent continuous layer due to reduced pressure during the homogenization process. The analytical and numerical results of applying the partially-homogenized-plasma EOS are consistent with experimental data on slowing down of laser-initiated shock wave in porous substance in comparison with continuous substance, and with experimental data on maintaining the quasi-consolidated state of a porous layer heated by X-ray pulse.

The calculation of homogenization degree is carried out in the approximation of the collision of plane flows of matter inside the pores. The use of this approximation is justified for the substances under consideration with a large porosity exceeding values of the order of 0.99, at which the particles of pore's walls fly within a pore the distances of 5–10 times greater than the initial wall's thickness. With a decrease in porosity, homogenization rate becomes sensitive to the geometry of initial porous structure, which, apparently, can be correctly taken into account only with using a numerical calculation.